\documentclass[12pt]{article}

\usepackage{amsmath,amssymb,amsbsy,amsfonts,amsthm,latexsym,
                        amsopn,amstext,amsxtra,euscript,amscd,color}
\usepackage{color,xcolor}
\usepackage{latexsym}
\usepackage{cite}
\usepackage{amsfonts}
\usepackage{amsfonts,mathrsfs}
\usepackage{graphicx}
\usepackage{psfrag}
\usepackage{subfigure}
\usepackage{url}
\usepackage{stfloats}
\usepackage{amsmath}

\newtheorem{thm}{Theorem}
\newtheorem{lem}{Lemma}

\newcommand{\quash}[1]{}

\begin{document}

\title{Linear complexity of Legendre-polynomial quotients}
\author{Zhixiong Chen\\
Provincial Key Laboratory of Applied Mathematics,\\ Putian University, Putian, Fujian 351100, P.R. China}

\maketitle

\begin{abstract}
We continue to investigate binary sequence $(f_u)$ over $\{0,1\}$ defined by $(-1)^{f_u}=\left(\frac{(u^w-u^{wp})/p}{p}\right)$
for integers $u\ge 0$, where $\left(\frac{\cdot}{p}\right)$ is the Legendre symbol and we restrict $\left(\frac{0}{p}\right)=1$. 
In an earlier work, the linear complexity of $(f_u)$ was determined for $w=p-1$ under the assumption of
 $2^{p-1}\not\equiv 1 \pmod {p^2}$. In this work, 
we give possible values on the linear complexity of $(f_u)$ for all $1\le w<p-1$ under the same conditions. We also state that the case of larger $w(\geq p)$
can be reduced to that of $0\leq w\leq p-1$. 
\end{abstract}



\section{Introduction}

For an odd prime $p$ and an integer $u$ with $\gcd(u,p)=1$, the {\it
Fermat quotient $q_p(u)$ \/} is defined as the unique integer
$$
q_p(u) \equiv \frac{u^{p-1} -1}{p} \bmod p ~~  \mathrm{with}~~ 0 \le
q_p(u) \le p-1,
$$
and
$$
q_p(kp) = 0, \qquad k \in \mathbb{Z}.
$$
An equivalent definition  is
$$
q_p(u)\equiv \frac{u^{p-1}-u^{p(p-1)}}{p}\bmod p.
$$
For all positive integers $w$, Chen and Winterhof extended the equation above
to define
$$
q_{p,w}(u)\equiv \frac{u^w-u^{wp}}{p} \bmod p ~~  \mathrm{with}~~ 0
\le q_{p,w}(u) \le p-1, ~~u\ge 0,
$$
which is called a \emph{polynomial quotient} in \cite{CW-Fq10}. In
fact $q_{p,p-1}(u)=q_{p}(u)$. We have the following relation between
$q_{p,w}(u)$ and $q_{p}(u)$:
\begin{equation}\label{poly-Fermat:relation}
q_{p,w}(u)\equiv \left\{
\begin{array}{ll}
-u^wwq_{p}(u) \bmod p, & \mathrm{if} ~ \gcd(u,p)=1,\\
0, &  \mathrm{if} ~ p|u ~\mathrm{and}~ w\ge 2,\\
k\bmod p, &  \mathrm{if} ~  u\equiv kp \pmod {p^2} ~\mathrm{and}~ w=1.
\end{array}
\right.
\end{equation}

Many number theoretic and cryptographic questions have been studied for polynomial quotients
\cite{AW-CC2011,Chen-CSIS2014,CD-DCC2013,CG-SETA2012,CNW-arxiv2013,COW-WAIFI2010,
CW-Fq10,CW-AA,DKC-IPL2012,GW-PMH2012,OS-SIAM2011}.

For any positive $w$ with $p|w$, we have $q_{p,w}(u)=0$ for all $u\ge 0$ by (\ref{poly-Fermat:relation}).
For $w$ with $p\nmid w$, write $w=w_1+w_2(p-1)$ with
$1\le w_1\le p-1$ and $w_2\ge 0$. By (\ref{poly-Fermat:relation})
again one can get for $\gcd(u,p)=1$
$$
q_{p,w_1+w_2(p-1)}(u)\equiv -u^{w_1}(w_1-w_2)q_{p}(u)\equiv
w_1^{-1}(w_1-w_2)q_{p,w_1}(u) \bmod p,
$$
which implies that $q_{p,w}$ for $w>p$ can be reduced to a polynomial quotient with $w<p$.
Hence, we may restrict ourselves to
$1\le w\le p-1$ from now on.

Two families of binary sequences have been considered in the literature. The first one is $(e_u)$  defined by
$$
e_u=\left\{
\begin{array}{ll}
0, & \mathrm{if}\,\ 0\leq q_{p,w}(u)/p< \frac{1}{2},\\
1, & \mathrm{if}\,\ \frac{1}{2}\leq q_{p,w}(u)/p< 1,
\end{array}
\right. \quad  u \ge 0.
$$
The second one  is $(f_u)$   defined  by
\begin{equation}\label{binarylegendre}
f_u=\left\{
\begin{array}{ll}
0, & \mathrm{if}\,\ \left(\frac{q_{p,w}(u)}{p}\right)=1\,\ \mathrm{or}\,\ q_{p,w}(u)=0, \\
1, & \mathrm{otherwise},
\end{array}
\right. \quad u \ge 0,
\end{equation}
where $\left( \frac{\cdot}{p} \right)$ is the Legendre symbol.

Both $(e_u)$ and $(f_u)$ have interesting cryptographic properties. In particular, when $2^{p-1}\not\equiv 1 \pmod {p^2}$,
the \emph{linear complexity} (see below for the notion) of $(e_u)$ has been determined if $w\in \{(p-1)/2,p-1\}$ in \cite{CG-SETA2012,CD-DCC2013}
and of $(f_u)$ if $w=p-1$ in \cite{Chen-CSIS2014}. While in \cite{CNW-arxiv2013}, the \emph{$k$-error linear complexity} of $(e_u)$ and $(f_u)$  has been determined for any $w$
under the assumption of 2 being a primitive root modulo $p^2$.
We note that ,if $2$ is a primitive root modulo $p^2$, then we
always have $2^{p-1}\not\equiv 1 \pmod {p^2}$.  But the converse is
not true, because there do exist such primes $p$, e.g., $p=43$. On the other hand, the primes $p$
satisfying $2^{p-1}\equiv 1 \pmod {p^2}$ are very rare. To date the
only known such primes are $p=1093$ and $p=3511$ and it was reported
that there are no new such primes $p< 4\times 10^{12}$, see
\cite{CDP1997}.

In this work, we will continue to investigate the linear complexity of $(f_u)$ for all $1\le w<p-1$ under the assumption of
$2^{p-1}\not\equiv 1 \pmod {p^2}$. Equation (\ref{poly-Fermat:relation}) is a key tool for our purpose. But, it seems that the technique here does not work  for $(e_u)$  when $w\not\in \{(p-1)/2,p-1\}$.

 We finally recall that the \emph{linear
complexity} $L((s_u))$ of a $T$-periodic sequence $(s_u)$ over the
binary field $\mathbb{F}_2$ is the least order $L$ of a linear
recurrence relation over $\mathbb{F}_2$
$$
s_{u+L} = c_{L-1}s_{u+L-1} +\cdots +c_1s_{u+1}+ c_0s_u\quad
\text{for}\,\ u \geq 0
$$
which is satisfied by $(s_u)$ and where $c_0=1, c_1, \ldots,
c_{L-1}\in \mathbb{F}_2.$ The polynomial
\begin{equation*}
M(x) = x^L + c_{L-1}x^{L-1} +\cdots+ c_0\in \mathbb{F}_2[x]
\end{equation*}
 is called the \emph{minimal
polynomial} of $(s_u)$. The \emph{generating polynomial} of $(s_u)$
is defined by
\begin{equation*}
s(x)=s_0+s_1x+s_2x^2+\cdots+s_{T-1}x^{T-1}\in \mathbb{F}_2[x].
\end{equation*}
It is easy to see that
$$
M(x)=(x^T-1)/\mathrm{gcd}\left(x^T-1,~ s(x)\right),
$$
hence
\begin{equation}
\label{eq:licom}
  L((s_u)) = T-\deg\left(\mathrm{gcd}(x^T-1, ~s(x))\right),
\end{equation}
which is the degree of the minimal polynomial,
see~\cite{CDR} for a more detailed exposition.

\section{Linear complexity}

From (\ref{poly-Fermat:relation}), we write
$$
H_{w}(u)\equiv \left\{
\begin{array}{ll}
-wq_{p}(u) \bmod p, & \mathrm{if} ~ \gcd(u,p)=1,\\
0, &  \mathrm{if} ~ p|u ~\mathrm{and}~ w\ge 2,\\
k\bmod p, &  \mathrm{if} ~  u\equiv kp\bmod {p^2} ~\mathrm{and}~ w=1.
\end{array}
\right.
$$

We will use the theory of cyclotomy, since $H_{w}: \mathbb{Z}_{p^2}^*\rightarrow \mathbb{Z}_p$ is a group homomorphism
\begin{equation}\label{H-homo}
H_{w}(uv) \equiv H_{w}(u) + H_{w}(v) \pmod p, ~~\gcd(uv,p)=1.
\end{equation}
by the fact that $q_p(uv) \equiv q_p(u) + q_p(v) \pmod p$ for $\gcd(uv,p)=1$, see e.g. \cite{OS-SIAM2011}.
In the context, we denote by
$\mathbb{Z}_{p}=\mathbb{F}_p=\{0,1,\ldots, p-1\}$ ~(respectively
$\mathbb{Z}_{p^2}=\{0,1,\ldots, p^2-1\}$) the residue class ring
modulo $p$ ~(respectively $p^2$) and by $\mathbb{Z}_{p^2}^*$ the
unit group of $\mathbb{Z}_{p^2}$.

Define
$$
D_l=\{u: 0\le u< p^2,~ \gcd(u,p)=1,~ H_{w}(u)\equiv l \pmod p\}
$$
for $l=0,1,\ldots,p-1$. Indeed, if $g$ is a (fixed) primitive root modulo $p^2$, we have by (\ref{H-homo})
$$
D_0=\{g^{kp} \bmod {p^2}: 0\le k<p\}
$$
and for $l=0,1,\ldots,p-1$, there exists an integer $0\le l_0<p$ such that $D_l=g^{l_0}D_0$. Hence each $D_l$ has the cardinality $|D_l|=p-1$. Naturally $D_0,D_1, \ldots, D_{p-1}$ form a partition of $\mathbb{Z}_{p^2}^*$.
Let $P=\{kp: 0\le k< p\}$.

We use the notation $aD_l=\{ab \pmod {p^2}: b\in
D_l\}$. Using \eqref{H-homo} we
have the following basic fact
\begin{enumerate}
\item[ (I).] $ aD_{l}= D_{l+l' \pmod p}$ if $a\in D_{l'}$.
 \end{enumerate}
Define
$$
 D_l(x)= \sum\limits_{u\in D_l}x^u \in \mathbb{F}_2[x],
 $$
for $l\in\{0,\ldots, p-1\}.$

\subsection{The case of even $w : 2\le w< p$.}

\begin{thm}
\label{thm:LC-even} Let $(f_u)$ be the $p^{2}$-periodic binary sequence
defined  in \eqref{binarylegendre} with even $w : 2\le w< p$.  Assume
that $2^{p-1}\not\equiv 1 \pmod {p^2}$ then,
\[
L((f_u))=\left\{
\begin{array}{ll}
p^2-p, & \mathrm{if}\,\ p \equiv 1 \pmod 4, \\
p^2-1, & \mathrm{if}\,\ p \equiv 3 \pmod 4.
\end{array}
\right.
\]
\end{thm}
Proof.
For even $w$, we have by (\ref{poly-Fermat:relation})
$$
(-1)^{f_{u}}=\left\{
\begin{array}{ll}
\left(\frac{H_{w}(u)}{p}\right), & \mathrm{if}~ (u \bmod {p^2})\in\mathbb{Z}_{p^2}^*, ~ \gcd(H_{w}(u),p)=1, \\
1, & \mathrm{otherwise},
\end{array}
\right.
$$
and define
$(f_u)$ in an equivalent way by
$$
f_u=\left\{
\begin{array}{ll}
0, & \mathrm{if}\,\ (u \bmod {p^2})\in \cup_{l\in \mathcal{Q}} D_l  \cup D_0\cup P,\\
1, & \mathrm{if}\,\ (u \bmod {p^2})\in \cup_{l\in \mathcal{N}} D_{l},
\end{array}
\right.
$$
here and hereafter $\mathcal{Q}$ denotes the set of quadratic residues modulo $p$ and $\mathcal{N}$ denotes the set of quadratic non-residues modulo $p$. We note that
the cardinality $|\mathcal{Q}|=|\mathcal{N}|=(p-1)/2$.

Then the proof of \cite[Th. 4]{Chen-CSIS2014} can help us to get the desired result. Here we present a detailed proof
for completeness.

Let
$$
\Lambda_{\ell}(x)=
\sum\limits_{l\in \mathcal{N}}D_{l+\ell}(x)\in
\mathbb{F}_2[x], ~~~\ell=0,\ldots,p-1.
$$
Clearly $\Lambda_0(x)$ is the generating polynomial of $(f_u)$. For $0\le \ell<p$, we only need to show $\Lambda_{\ell}(\beta)\neq 0$  for  a  primitive $p^2$-th root of unity $\beta \in \overline{\mathbb{F}}_{2}$ by  (\ref{eq:licom}).

We assume that $\Lambda_{\ell_0}(\beta)=0$ for some $0\le \ell_0<p$. Since $2^{p-1}\not\equiv 1 \pmod {p^2}$, we set $H_{w}(2)=-w2^{w}q_p(2)=\mu\neq 0$. Then by Fact (I)  we have for $0\le j<p$
$$
0=\left(\Lambda_{\ell_0}(\beta)\right)^{2^j}=\Lambda_{\ell_0}(\beta^{2^j})=\Lambda_{\ell_0+j\mu}(\beta),
$$
where the subscript of $\Lambda$ is reduced modulo $p$. That is, $\Lambda_{\ell}(\beta)=0$ for all $0\le \ell<p$.
We have furtherly $\Lambda_{\ell}(\beta^u)=0$  by Fact (I) again for all $u\in \mathbb{Z}_{p^2}^*$ and $0\le \ell<p$.
On the other hand, all ($p^2-p$ many) elements $\beta^u$ for $u\in \mathbb{Z}_{p^2}^*$ are roots of
$$
\Phi(x)=1+x^p+x^{2p}+\ldots+x^{(p-1)p}\in \mathbb{F}_2[x],
$$
which has no other roots. Hence  we have
$$
\Phi(x)|\Lambda_0(x) ~~ \mathrm{in}~ \overline{\mathbb{F}}_2[x].
$$
 Let
\begin{equation}\label{pi}
\Lambda_0(x)\equiv \Phi(x)\pi(x) \pmod {x^{p^2}-1}.
\end{equation}
Using the fact that
$$
x^p\Phi(x)  \equiv \Phi(x) \pmod {x^{p^2}-1},
$$
we restrict $\deg(\pi(x))<p$. However, $\Lambda_0(x)$ has $(p-1)^2/2$ terms and the
right hand side of (\ref{pi}) has $pt$ terms if $\pi(x)$ has $t$ terms, a
contradiction.  So we conclude that
$\Lambda_{\ell}(\beta^u)\neq 0$  for all $u\in\mathbb{Z}_{p^2}^*$ and $0\le \ell<p$.

On the other hand,   we have
$$
\Lambda_{\ell}(\beta^{kp})=\left\{
\begin{array}{ll}
0, & \mathrm{if}~ k= 0,\\
(p-1)/2, & \mathrm{if}~ 1\le k<p,
\end{array}
\right.
$$
for $0\le \ell<p$. We draw a conclusion that $\Lambda_0(x)$, the generating polynomial of $(f_u)$, and $x^{p^2}-1$ have exactly $p$ many common roots $\beta^{kp}~(0\le k<p)$ if $p \equiv 1 \pmod 4$ and one common root $\beta^0$ otherwise. ~\hfill $\square$

\subsection{The case of odd $w : 3\le w<p$.}

For $l\in\{0,\ldots, p-1\}$, we define
\begin{equation*}
Q_l=\left\{u\in D_l:
\left(\frac{u}{p}\right)=1\right\}\quad\text{and}\quad
N_l=\left\{u\in D_l: \left(\frac{u}{p}\right)=-1\right\}.
\end{equation*}
  Using \eqref{H-homo} aagain we
have the following facts:
\begin{enumerate}
\item[(II).] $ aQ_{l}=Q_{l+l' \pmod p} $ if $a\in Q_{l'}$.
\item[(III).] $ aN_{l} = N_{l+l' \pmod p}$ if $a\in   Q_{l'}$.
\item[(IV).] $ aQ_{l}= N_{l+l' \pmod p} $ if $a\in N_{l'}$.
\item[(V).] $ aN_{l}=Q_{l+l' \pmod p} $ if $a\in
  N_{l'}$.
\item[(VI).] $\{u\pmod p: u\in D_l\}=\{1,2,\ldots, p-1\},~ l\in\{0,1,\ldots, p-1\}$.
             Hence, $|Q_l|=|N_l|=(p-1)/2$.
\end{enumerate}
Define
\begin{equation*}
 Q_l(x)= \sum\limits_{u\in Q_l}x^u \in \mathbb{F}_2[x],\quad
 N_l(x)= \sum\limits_{u\in N_l}x^u \in \mathbb{F}_2[x]
\end{equation*}
for $l\in\{0,\ldots, p-1\}$. We see that $ D_l(x)=Q_l(x)+N_l(x)$. Now we present the results in the following two theorems.

\begin{thm}
\label{thm:LC-odd} Let $(f_u)$ be the $p^{2}$-periodic binary sequence
defined  in \eqref{binarylegendre} with odd $3\le w<p$.  Assume
that $2^{p-1}\not\equiv 1 \pmod {p^2}$ then,
\begin{equation*}
  L((f_u))=
  \begin{cases}
    p^2-p~~\mathrm{or}~~ (p^2-p)/2, & \mathrm{if}\,\ p \equiv 1 \pmod 8, \\
    p^2-1~~\mathrm{or}~~ (p^2+p)/2-1, & \mathrm{if}\,\ p \equiv -1 \pmod
    8,\\
    p^2-p, & \mathrm{if}\,\ p \equiv -3 \pmod 8, \\
    p^2-1, & \mathrm{if}\,\ p \equiv 3 \pmod 8.
  \end{cases}
\end{equation*}
\end{thm}
 Proof.
 For odd $w\ge 3$, we have by (\ref{poly-Fermat:relation})
$$
(-1)^{f_{u}}=\left\{
\begin{array}{ll}
\left(\frac{u}{p}\right)\left(\frac{H_{w}(u)}{p}\right), & \mathrm{if}~ (u \bmod {p^2})\in\mathbb{Z}_{p^2}^*, ~ \gcd(H_{w}(u),p)=1, \\
1, & \mathrm{otherwise},
\end{array}
\right.
$$
and define $(f_u)$ in an equivalent way by
 $$
f_u=\left\{
\begin{array}{ll}
0, & \mathrm{if}\,\  (u \bmod {p^2})\in \cup_{l\in \mathcal{Q}} Q_l \cup \cup_{l\in \mathcal{N}} N_{l} \cup D_0\cup P,\\
1, & \mathrm{if}\,\  (u \bmod {p^2})\in \cup_{l\in \mathcal{Q}} N_l \cup \cup_{l\in \mathcal{N}} Q_{l}.
\end{array}
\right.
$$

Then the generating polynomial of $(f_u)$ is
$$
\begin{array}{rl}
G^{\mathrm{odd}}(x)= \sum\limits_{u=0}^{p^2-1}e_ux^u= & \sum\limits_{l\in \mathcal{Q}}N_l(x)+
\sum\limits_{l\in \mathcal{N}}Q_l(x) \\
=& \sum\limits_{l\in \mathcal{N}}D_l(x)+
\sum\limits_{l=1}^{p-1}N_l(x)
\in \mathbb{F}_2[x].
\end{array}
$$
Below we will consider the common roots of $G^{\mathrm{odd}}(x)$ and $x^{p^2}-1$.
The number of  the common roots will lead to the values of  linear
complexity of $(f_u)$ by \eqref{eq:licom}. We need the following lemmas, which can be proved by following the way of \cite{CG-SETA2012}.

\begin{lem}\label{lem:lemma-kp}\cite{CG-SETA2012}
Let $\beta \in \overline{\mathbb{F}}_{2}$ be  a  primitive $p^2$-th root of
unity. We have
$$
G^{\mathrm{odd}}(\beta^n)= \left \{
\begin{array}{ll}
0, & \mathrm{if}\,\ n=0,\\
\frac{p-1}{2}, & \mathrm{if}\,\ n=kp,~k=1,\ldots,p-1.
\end{array}
\right.
$$
\end{lem}

\begin{lem}\label{lem:lemma-allN}\cite{CG-SETA2012}
Let $\beta \in \overline{\mathbb{F}}_{2}$ be  a  primitive $p^2$-th root of
unity. For all $n\in \mathbb{Z}_{p^2}^*$, we have
$\sum\limits_{l=0}^{p-1}N_l(\beta^n)=0$.
\end{lem}

\begin{lem}\label{lem:lemma-allD}\cite{CG-SETA2012}
Let $\beta \in \overline{\mathbb{F}}_{2}$ be  a primitive $p^2$-th root of
unity. If  $2\in D_{\ell_0}$ for some $1\le \ell_0\le p-1$, we have
$D_l(\beta^n)\not =0$ for all $0\le l\le p-1$ and $n \in
\mathbb{Z}_{p^2}^*$.
\end{lem}

We use Lemmas \ref{lem:lemma-allN} and \ref{lem:lemma-allD} to show the following lemma.

\begin{lem}\label{lem:lemma-e}\cite{CG-SETA2012}
Let $\beta \in \overline{\mathbb{F}}_{2}$ be  a primitive $p^2$-th root of
unity, then
\begin{enumerate}
\item[(1).] If $2\in Q_{\ell_0}$ for some $1\le \ell_0\le p-1$, then there
  exist(s) either exactly $(p^2-p)/2$ many or no $n \in \mathbb{Z}_{p^2}^*$ such that
  $G^{\mathrm{odd}}(\beta^{n}) =0$.
\item[(2).] If $2\in N_{\ell_0}$ for some $1\le \ell_0\le p-1$, then
  $G^{\mathrm{odd}}(\beta^{n})\neq 0$ for all $n\in \mathbb{Z}_{p^2}^*$.
\end{enumerate}
\end{lem}

Now we continue the proof of  Theorem~\ref{thm:LC-odd}.

First we suppose that $\left(\frac{2}{p}\right)=1$. In this case,
$p\equiv \pm 1\pmod 8$. If $p \equiv 1 \pmod 8$, we have
$G^{\mathrm{odd}}(\beta^n)=0$ if $n\in \{kp: 0\le k\le p-1\}$ by Lemma
\ref{lem:lemma-kp} and there are either no numbers in
$\mathbb{Z}_{p^2}^*$ or $p(p-1)/2$ many $n \in \mathbb{Z}_{p^2}^*$
such that $G^{\mathrm{odd}}(\beta^{n}) =0$ by Lemma \ref{lem:lemma-e}. Then the number of
the common roots of $G^{\mathrm{odd}}(x)$ and $x^{p^2}-1$ is either $p$ or
$(p^2+p)/2$ and hence the linear complexity of $(f_u)$ is $p^2-p$ or
$(p^2-p)/2$. For the case of $p \equiv -1 \pmod 8$, the result follows
similarly.

Under the condition of $\left(\frac{2}{p}\right)=-1$, it can be proved
in a similar way. ~\hfill $\square$

We calculate the linear complexity of $(f_u)$ for all primes $p<200$. We list some examples in Table 1. The experiment results illuminate that, when $p\equiv -1 \bmod 8$ the linear complexity only equals $\frac{p^2+p}{2}-1$.
So we might ask whether there exists $p$ such
that linear complexity equals $p^2-1$ if $p\equiv -1 \bmod 8$.

\begin{center}
\begin{tabular}{cccccccc}
\hline\noalign{\smallskip}
$p$ & mod 8 & linear complexity  &~ remark \\
 \noalign{\smallskip} \hline \noalign{\smallskip}
$41$  &  1    & $820=\frac{p^2-p}{2}$   &    \\
$73$  &  1    &  $5256=p^2-p$               &    \\
$11$  &  3    &  $120=p^2-1$                & satisfying \cite[Th.2]{DKC-IPL2012}  \\
$43$  &  3    &  $1848=p^2-1$               &   \\
$13$  &  $-3$  &  $156=p^2-p$                 & satisfying \cite[Th.2]{DKC-IPL2012}\\
$109$  & $-3$  &  $11772=p^2-p $             & \\
$23$  &   $-1$  &  $275=\frac{p^2+p}{2}-1$     &   \\
  \hline
\end{tabular}
\\ Table 1. Linear complexity of $(f_u)$ for odd $3\le w<p$.
\end{center}

\subsection{The case of $w=1$.}

\begin{thm}
\label{thm:LC-one} Let $(f_u)$ be the $p^{2}$-periodic binary sequence
defined  in \eqref{binarylegendre} with $w=1$.  Assume
that $2^{p-1}\not\equiv 1 \pmod {p^2}$ then,
\begin{equation*}
  L((f_u))=
  \begin{cases}
    p^2-p~~\mathrm{or}~~ (p^2-p)/2, & \mathrm{if}\,\ p \equiv 1 \pmod 4, \\
    p^2-p+1~~\mathrm{or}~~ (p^2-p)/2+1, & \mathrm{if}\,\ p \equiv 3 \pmod 4.
  \end{cases}
\end{equation*}
\end{thm}
 Proof. We have by (\ref{poly-Fermat:relation})
$$
(-1)^{f_{u}}=\left\{
\begin{array}{ll}
\left(\frac{u}{p}\right)\left(\frac{H_{1}(u)}{p}\right), & \mathrm{if}~ (u \bmod {p^2})\in\mathbb{Z}_{p^2}^*,~ \gcd(H_{1}(u),p)=1, \\
\left(\frac{k}{p}\right),                                   & \mathrm{if}~ u\equiv kp \bmod {p^2}, ~1\le k<p, \\
1, & \mathrm{otherwise},
\end{array}
\right.
$$
and define $(f_u)$ in an equivalent way by
$$
f_u=\left\{
\begin{array}{ll}
0, & \mathrm{if}\,\ u\in \cup_{l\in \mathcal{Q}} Q_l \cup \cup_{l\in \mathcal{N}} N_{l} \cup D_0\cup \{0\} \cup\{lp :  l\in \mathcal{Q}\},\\
1, & \mathrm{if}\,\ u\in \cup_{l\in \mathcal{Q}} N_l \cup \cup_{l\in \mathcal{N}} Q_{l}\cup \{lp :  l\in \mathcal{N}\}.
\end{array}
\right.
$$
Then the generating polynomial $G^{\mathrm{one}}(x)$ of $(f_u)$ is
$$
\begin{array}{rl}
G^{\mathrm{one}}(x)= \sum\limits_{u=0}^{p^2-1}e_ux^u= & \sum\limits_{l\in \mathcal{Q}}N_l(x)+
\sum\limits_{l\in \mathcal{N}}Q_l(x)+
\sum\limits_{l\in \mathcal{N}}x^{lp} \\
=& \sum\limits_{l\in \mathcal{N}}D_l(x)+
\sum\limits_{l=1}^{p-1}N_l(x)+
\sum\limits_{l\in \mathcal{N}}x^{lp}
\in \mathbb{F}_2[x].
\end{array}
$$
We need to compute the number of $n\in\mathbb{Z}_{p^2}$ with $G^{\mathrm{one}}(\beta^n)=0$ for a primitive $p^2$-th root of
unity $\beta \in \overline{\mathbb{F}}_{2}$.

\begin{lem}\label{lem:lemma-kp-one}
Let $\beta \in \overline{\mathbb{F}}_{2}$ be  a  primitive $p^2$-th root of
unity. We have
$$
G^{\mathrm{one}}(\beta^n)= \left \{
\begin{array}{ll}
\frac{p-1}{2}, & \mathrm{if}\,\ n=0,\\
0, & \mathrm{if}\,\ n=kp,~k=1,\ldots,p-1.
\end{array}
\right.
$$
\end{lem}

\begin{lem}\label{lem:lemma-allD-one}
Let $\beta \in \overline{\mathbb{F}}_{2}$ be  a primitive $p^2$-th root of
unity. If  $2\in D_{\ell_0}$ for some $1\le \ell_0\le p-1$, we have
$D_l(\beta^n)\not =1$ for all $0\le l\le p-1$ and $n \in
\mathbb{Z}_{p^2}^*$.
\end{lem}
 (Proof of Lemma \ref{lem:lemma-allD-one}). Let $\widetilde{D}_l(x)=1+D_l(x)$. Suppose that there is an $n_0\in D_{i_0}$ for
  some $0\le i_0\le p-1$ such that $D_{l_0}(\beta^{n_0})=1$ for some
  $0\le l_0\le p-1$. Then we have $\widetilde{D}_l(\beta^{n_0})=0$ and drive that each polynomial $\widetilde{D}_l(x)$ has at least $p(p-1)$ many roots using the proof of Lemma \ref{lem:lemma-allD}. We get a contradiction  since at least one
$\widetilde{D}_l(x)$ has degree $<p^2-p$. We complete the proof of Lemma \ref{lem:lemma-allD-one}.

\begin{lem}\label{lem:lemma-QN-one}
Let $\theta \in \overline{\mathbb{F}}_{2}$ be  a  primitive $p$-th root of
unity. Then
\begin{enumerate}
\item[(1).] we have $ \sum\limits_{j\in \mathcal{N}}\theta^{j}+\sum\limits_{j\in \mathcal{Q}}\theta^{j}=1$.
\item[(2).] we have
$$
\sum\limits_{j\in \mathcal{N}}\theta^{nj}=\left\{
            \begin{array}{ll}
               \sum\limits_{j\in \mathcal{N}}\theta^{j}, & \mathrm{if}\,\ n\in Q_l,\\
               \sum\limits_{j\in \mathcal{Q}}\theta^{j}, & \mathrm{if}\,\ n\in N_l,
\end{array}
\right.
$$
and
$$
\sum\limits_{j\in \mathcal{Q}}\theta^{nj}=\left\{
            \begin{array}{ll}
               \sum\limits_{j\in \mathcal{Q}}\theta^{j}, & \mathrm{if}\,\ n\in Q_l,\\
               \sum\limits_{j\in \mathcal{N}}\theta^{j}, & \mathrm{if}\,\ n\in N_l,
\end{array}
\right.
$$
for $0\le l<p$.
\end{enumerate}
\end{lem}

\begin{lem}\label{lem:lemma-e-one}
Let $\beta \in \overline{\mathbb{F}}_{2}$ be  a primitive $p^2$-th root of
unity.  If $2\in Q_{\ell_0}$ or  $2\in N_{\ell_0}$ for some $1\le \ell_0\le p-1$, then there
  exist(s) either exactly $(p^2-p)/2$ many or no $n \in \mathbb{Z}_{p^2}^*$ such that
  $G^{\mathrm{one}}(\beta^{n}) =0$.
\end{lem}
 (Proof of Lemma \ref{lem:lemma-e-one}). It is easy to see that for all $n \in \mathbb{Z}_{p^2}^*$
  \begin{equation*}
 G^{\mathrm{one}}(\beta^{n})=\sum\limits_{l\in \mathcal{N}}D_l(\beta^{n})+ N_0(\beta^n) + \sum\limits_{l\in \mathcal{N}}\beta^{nlp}.
  \end{equation*}
by Lemma~\ref{lem:lemma-allN}. Let $\xi=\sum\limits_{l\in \mathcal{N}}\beta^{lp}$  and as before
$$
\Lambda_{\ell}(x)=
\sum\limits_{l\in \mathcal{N}}D_{l+\ell}(x)\in
\mathbb{F}_2[x], ~~~\ell=0,\ldots,p-1.
$$
Then together with Facts (I)-(V), we have
\begin{equation}\label{eq:G-one}
G^{\mathrm{one}}(\beta^{n})=
  \begin{cases}
    \Delta_{l}(\beta)+N_l(\beta)+\xi,  &  \text{if } n \in Q_l,\\
    \Delta_{l}(\beta)+Q_l(\beta)+1+\xi, &  \text{if } n \in N_l,
  \end{cases}
\end{equation}
which indicates $G^{\mathrm{one}}(\beta^{m})\neq G^{\mathrm{one}}(\beta^{n})$ for $m\in Q_l$ and
$n\in N_l$ by Lemma \ref{lem:lemma-allD-one}.

Case (i). Let $2\in Q_{\ell_0}$. We suppose that $G^{\mathrm{one}}(\beta^{n_0})=0$ for some $n_0\in D_{i_0}, ~ 0\le i_0\le p-1$. If $n_0\in Q_{i_0}$, then $2^jn_0\in
Q_{j\ell_0+i_0\pmod p}$ for $0\le j\le p-1$. For all $n\in Q_{j\ell_0+i_0\pmod p}$, by (\ref{eq:G-one}) we derive
\begin{eqnarray*}
  G^{\mathrm{one}}(\beta^n)& =& \Delta_{{j\ell_0+i_0\pmod
      p}}(\beta)+N_{{j\ell_0+i_0\pmod
      p}}(\beta)+\xi\\
  &=& G^{\mathrm{one}}(\beta^{2^{j}n_0})=(G^{\mathrm{one}}(\beta^{n_0}))^{2^j}=0,
\end{eqnarray*}
which also implies  $ G^{\mathrm{one}}(\beta^n)\neq 0 $
for all $n\in N_{j\ell_0+i_0\pmod p}$. So we have for $n \in
\mathbb{Z}_{p^2}^*$
$$
G^{\mathrm{one}}(\beta^n)=0~~~\mathrm{iff}~~ n\in Q_0\cup Q_1\cup \cdots \cup
Q_{p-1}.
$$
Similarly, if $n_0\in N_{i_0}$, we have
$$
G^{\mathrm{one}}(\beta^n)=0~~~\mathrm{iff}~~ n\in N_0\cup N_1\cup \cdots \cup
N_{p-1}.
$$

Case (ii). For the case of $2\in N_{\ell_0}$, i.e.,
$\left(\frac{2}{p}\right)=-1$, if $G^{\mathrm{one}}(\beta^{n_0}) =0$ for some $n_0\in
Q_{i_0}$, then we have $2^jn_0\in N_{j\ell_0+i_0\pmod p}$ for odd $0\le j\le 2p-1$ and $2^jn_0\in Q_{j\ell_0+i_0\pmod p}$ for even $0\le j\le 2p-1$.

We note that $\xi^2=1+\xi$ by Lemma \ref{lem:lemma-QN-one}. So for even $j$ we have by (\ref{eq:G-one})
\begin{eqnarray*}
G^{\mathrm{one}}(\beta^{2^{j}n_0})& = &\Delta_{{j\ell_0+i_0\pmod p}}(\beta)+N_{{j\ell_0+i_0\pmod p}}(\beta)+\xi \\
&=& \Delta_{0}(\beta^{n_0})^{2^j}+N_{0}(\beta^{n_0})^{2^j}+\xi^{2^j} \\
&=&(G^{\mathrm{one}}(\beta^{n_0}))^{2^j}=0
\end{eqnarray*}
and so $G^{\mathrm{one}}(\beta^n)=0$ iff $ n\in Q_0\cup Q_1\cup \cdots \cup
Q_{p-1}$.

Similarly, if $G^{\mathrm{one}}(\beta^{n_0}) =0$ for some $n_0\in
N_{i_0}$, we will get $G^{\mathrm{one}}(\beta^n)=0$ iff $ n\in N_0\cup N_1\cup \cdots \cup
N_{p-1}$.

Thus we conclude that there exist $p(p-1)/2$ many $n \in
\mathbb{Z}_{p^2}^*$ such that $G^{\mathrm{one}}(\beta^{n}) =0$ since both $Q_l$ and
$N_l$ contain $(p-1)/2$ elements. We complete the proof of Lemma \ref{lem:lemma-e-one}.

Finally we finish the proof of  Theorem~\ref{thm:LC-one} by Lemmas \ref{lem:lemma-kp-one} and \ref{lem:lemma-e-one}.
 ~\hfill $\square$

We calculate the linear complexity of $(f_u)$ for all primes $p<200$ and list some examples in Table 2. The experiment results illuminate that, no primes $p$ such that the linear complexity equals $p^2-p$, $\frac{p^2-p}{2}+1$, $p^2-p+1$ when $p\equiv 1,3,7 \bmod 8$, respectively. We leave it open.

\begin{center}
\begin{tabular}{cccccccc}
\hline\noalign{\smallskip}
$p$ & mod 8 & linear complexity  &~ remark \\
 \noalign{\smallskip} \hline \noalign{\smallskip}
$17$  &  1    & $136=\frac{p^2-p}{2}$   &    \\
$11$  &  3    &  $111=p^2-p+1$                & satisfying \cite[Th.2]{DKC-IPL2012}  \\
$43$  &  3    &  $1807=p^2-p+1$               &   \\
$13$  &  $-3$ &  $156=p^2-p$                 & satisfying \cite[Th.2]{DKC-IPL2012}\\
$109$  & $-3$   &  $11772=p^2-p $             & \\
$157$  & $-3$  & $12246=\frac{p^2-p}{2}$   &    \\
$23$  &   $-1$  &  $254=\frac{p^2-p}{2}+1$     &   \\
  \hline
\end{tabular}
\\ Table 2. Linear complexity of $(f_u)$ for $w=1$.
\end{center}

\section{Final remarks}

In this work, we have determined all possible values on the linear complexity of the binary sequences defined via computing the  Legendre symbol of polynomial quotients. 
The achievement extends corresponding results of the Legendre-Fermat quotients studied in \cite{Chen-CSIS2014}.

It is interesting to consider the linear complexity of $(e_u)$ defined in Section 1 for $w\not\in\{p-1,(p-1)/2\}$.

\section*{Acknowledgements}
The work was partially supported  by the National Natural Science
Foundation of China under grant No. 61373140.

\end{document}